\documentstyle[amssymb]{article}

\newcommand{\naturals}{{\Bbb N}}

\providecommand{\rmd}{\mathrm{d}}

\makeatletter
\def\@cite#1#2{$^{\mathrm{({#1\if@tempswa , #2\fi})}}$}
\makeatother

\newcommand{\htmladdnormallink}[2]{#1}

\begin{document}

\title{Why Occam's Razor}

\author{
{\bf Russell K. Standish}\\
{\em School of Mathematics}\\
{\em University of New South Wales}\\
{\em Sydney, 2052, Australia}\\
{\em R.Standish@unsw.edu.au, http://parallel.hpc.unsw.edu.au/rks}
}

\maketitle

\begin{abstract}
Ensemble theories have received a lot of interest recently as a
  means of explaining a lot of the detailed complexity observed in
  reality by a vastly simpler description ``every possibility exists''
  and a selection principle ({\em Anthropic Principle}) ``we only
  observe that which is consistent with our existence''.  In this
  paper I show why, in an ensemble theory of the universe, we should
  be inhabiting one of the elements of that ensemble with least
  information content that satisfies the anthropic principle. This
  explains the effectiveness of aesthetic principles such as Occam's
  razor in predicting usefulness of scientific theories. I also show,
  with a couple of reasonable assumptions about the phenomenon of
  consciousness, the linear structure of quantum mechanics can be
  derived.  
\end{abstract}

\maketitle

\section{INTRODUCTION}

Wigner\cite{Wigner67} once remarked on ``the unreasonable
effectiveness of mathematics'', encapsulating in one phrase the
mystery of why the scientific enterprise is so successful. There is an
aesthetic principle at large, whereby scientific theories are chosen
according to their beauty, or simplicity. These then must be tested by
experiment --- the surprising thing is that the aesthetic quality of a
theory is often a good predictor of that theory's explanatory and
predictive power. This situation is summed up by William of Ockham,
``Entities should not be multiplied unnecessarily'', known as Occam's
Razor.

We start our search into an explanation of this mystery with the
{\em anthropic principle}\cite{Barrow-Tipler86}. This is normally cast
into either a weak form (that physical reality must be consistent
with our existence as conscious, self-aware entities) or a strong
form (that physical reality is the way it is {\em because} of our
existence as conscious, self-aware entities). The anthropic
principle is remarkable in that it generates significant constraints
on the form of the
universe\cite{Barrow-Tipler86,Tegmark98}. The two main
explanations for this are the {\em Divine Creator explanation} (the
universe was created deliberately by God to have properties
sufficient to support intelligent life), or the {\em Ensemble
explanation}\cite{Tegmark98} (that there is a set, or ensemble, of
different universes, differing in details such as physical
parameters, constants and even laws, however, we are only aware of
such universes that are consistent with our existence). In the
Ensemble explanation, the strong and weak formulations of the
anthropic principle are equivalent.

Tegmark introduces an ensemble theory based on the idea that every
self-consistent mathematical structure be accorded the ontological
status of {\em physical existence}. He then goes on to categorize
mathematical structures that have been discovered thus far (by
humans), and argues that this set should be largely universal, in
that all self-aware entities should be able to uncover at least the
most basic of these mathematical structures, and that it is unlikely
we have overlooked any equally basic mathematical structures.

An alternative ensemble approach is that of
Schmidhuber's\cite{Schmidhuber97} --- the ``Great Programmer''. This
states that all possible halting programs of a universal Turing
machine have physical existence. Some of these programs' outputs will
contain self-aware substructures --- these are the programs deemed
interesting by the anthropic principle. Note that there is no need for
the UTM to actually exist, nor is there any need to specify which UTM
is to be used --- a program that is meaningful on UTM$_1$ can be
executed on UTM$_2$ by prepending it with another program that
describes UTM$_1$ in terms of UTM$_2$'s instructions, then executing
the individual program. Since the set of halting programs (finite
length bitstrings) is isomorphic to the set of whole numbers
$\naturals$, an enumeration of $\naturals$ is sufficient to generate
the ensemble that contains our universe.  In a later
paper\cite{Schmidhuber00}, Schmidhuber extends his ensemble to
non-halting programs, and consider the consequences of assuming that
this ensemble is generated by a machine with bounded resources.

 Each self-consistent mathematical structure (member of
the Tegmark ensemble) is completely described by a finite set of
symbols, and a countable set of axioms encoded in those symbols, and a
set of rules (logic) describing how one mathematical statement may be
converted into another.\footnote{Strictly speaking, these systems are
  called recursively enumerable formal systems, and are only a subset
  of the totality of mathematics, however this seem in keeping with
  the spirit of Tegmark's suggestion} These axioms may be encoded as a
bitstring, and the rules encoded as a program of a UTM that enumerates
all possible theorems derived from the axioms, so each member of the
Tegmark ensemble may be mapped onto a Schmidhuber one.\footnote{In the
  case of an infinite number of axioms, the theorems must be
  enumerated using a dovetailer algorithm. The dovetailer algorithm is
  a means of walking an infinite level tree, such that each level is
  visited in finite time. An example is that for a $n$-ary tree, the
  nodes on the $i$th level are visited between steps $n^i$ and
  $n^{i+1}-1$.}. The Tegmark ensemble must be contained within the
Schmidhuber one.

An alternative connection between the two ensembles is that the
Schmidhuber ensemble is a self-consistent mathematical structure, and
is therefore an element of the Tegmark one. However, all this implies
is that one element of the ensemble may in fact generate the complete
ensemble again, a point made by Schmidhuber in that the ``Great
Programmer'' exists many times, over and over in a recursive manner
within his ensemble. This is now clearly true also of the Tegmark ensemble.

\section{UNIVERSAL PRIOR}

In this paper, I adopt a Schmidhuber ensemble consisting of all
infinite length bitstrings, denoted $\{0,1\}^\infty$. I call these infinite
length strings {\em descriptions}. By contrast to Schmidhuber, I
assume a uniform measure over these descriptions --- no particular
string is more likely than any other. It can be shown that the
cardinality of $\{0,1\}^\infty$ is the same as the cardinality of the
reals, $c$. This set cannot be enumerated by a dovetailer algorithm,
rather the dovetailer algorithm enumerates all finite length prefixes
of these descriptions.  Whereas in Schmidhuber's
1997\cite{Schmidhuber97} paper, the existence of the dovetailer
algorithm explains the ease with which the ``Great Programmer'' can
generate the ensemble of universes, I merely assume the pre-existence
of all possible descriptions. The information content of this complete
set is precisely zero, as no bits are specified. It is ontologically
equivalent to Nothing. This has been called the ``zero information
principle''.

Since some of these descriptions describe self aware substructures, we
can ask the question of what these observers observe.  An observer
attaches sequences of {\em meanings} to sequences of prefixes of one
of these strings. A meaning belongs to a countable set, which may be
enumerated by the whole numbers. Thus the act of observation may
formalised as a map $O:[0,1]^\infty\rightarrow\naturals$. If $O(x)$ is a
{\em computable} (also known as a {\em recursive}) function, then
$O(x)$ is equivalent to a Turing machine, for which every input
halts. It is important to note that observers must be able to evaluate
$O(x)$ within a finite amount of subjective time, or the observer
simply ceases to be. The restriction to computable $O(x)$ connects
this viewpoint with the original viewpoint of Schmidhuber.

Another interpretation of this scenario is a state machine, possibly
finite, consuming bits of an infinite length string. As each bit is
consumed, the current state of the machine is the meaning attached to
the prefix read so far.


Under the mapping $O(x)$, some descriptions encode for identical
meanings as other descriptions, so one should equivalence class the
descriptions. In particular, strings where the bits after some bit
number $n$ are ``don't care'' bits, are in fact equivalence classes of
all strings that share the first $n$ bits in common. One can see that
the size of the equivalence class drops off exponentially with the
amount of information encoded by the string. Under $O(x)$, the amount
of information is not necessarily equal to the length of the string,
as some of the bits may be redundant. The sum
\begin{equation}\label{Universal Prior}
P_O(s)=\sum_{p:O(p)=s} 2^{-|p|},
\end{equation}
where $|p|$ means the number of bits of $p$ consumed by $O$ in
returning $s$, gives the size of the equivalence class of all descriptions
having meaning $s$ This measure distribution is known as a {\em
universal prior}, or alternatively a Solomonoff-Levin distribution, in
the case where $O(x)$ is a universal prefix Turing machine\cite{Li-Vitanyi97}.

The quantity
\begin{equation}
{\cal C}_O(x) = -\log_2 P_O(O(x))
\end{equation}
is a measure of the information content, or {\em complexity} of a
description $x$. If only the first $n$ bits of the string are
significant, with no redundancy, then it is easy to see ${\cal
C}_O(x)=n$. Moreover, if $O$ is a universal prefix Turing machine,
then the coding theorem\cite{Li-Vitanyi97} assures that ${\cal
C}(x)\approx K(x)$, where $K(x)$ is the usual Kolmogorov complexity,
up to a constant independent of the length of $x$.

If we assume the {\em self-sampling
assumption}\cite{Leslie96,Carter83}, essentially that we expect to
find ourselves in one of the universes with greatest measure, subject
to the constraints of the anthropic principle. This implies we should
find ourselves in one of the simplest (in terms of ${\cal C}_O$)
possible universes capable of supporting self-aware substructures
(SASes). This is the origin of physical law --- why we live in a
mathematical, as opposed to a magical universe. This is why aesthetic
principles, and Ockam's razor in particular are so successful at
predicting good scientific theories. This might also be called the
``minimum information principle''.

A final comment to highlight the distinction between this approach and
Schmidhuber's. Schmidhuber assumes that there is a given universal
Turing machine $U$ which generates the ensemble we find ourselves in. He
even uses the term ``Great Programmer'' to underscore
this. Ontologically, this is no more difficult than assuming there is
an ultimate theory of everything --- ie a final set of equations from
which all of physics can be derived. Occam's razor is a consequence of
the resource constraints of $U$. In my approach, there is no given
laws or global interpreter. By considering just the
resource constraints of the observer, even in the case of the
ensemble having a uniform measure, Occam's razor still applies.

\section{THE WHITE RABBIT PARADOX}

An important criticism leveled at ensemble theories is what John
Leslie calls {\em the failure of induction}\cite[\S4.69]{Leslie89}. If
all possible universes exist, then what is to say that our orderly,
well-behaved universe won't suddenly start to behave in a disordered
fashion, such that most inductive predictions would fail in them. This
problem has also been called the White Rabbit paradox\cite{Marchal95},
presumably in a literary reference to Lewis Carrol.

This sort of issue is addressed by consideration of measure. We should
not worry about the universe running off the rails, provided it is
extremely unlikely to do so. Note that Leslie uses the term {\em
  range} to mean what we mean by {\em measure}. At first
consideration, it would appear that there are vastly more ways for a
universe to act strangely, than for it to stay on the straight and
narrow, hence the paradox.

Evolution has taught us to be efficient classifiers of patterns, and
to be robust in the presence of errors. It is important to know the
difference between a lion and a lion-shaped rock, and to establish
that difference in real time. Only a finite number of the
description's bits are processed by the classifier, the remaining
being ``don't care'' bits.  Around each compact description is a cloud
of completely random descriptions considered equivalent by the
observer. The size of this cloud decreases exponentially with the
complexity of the description. 

This requirement imposes a significant condition on $O(x)$. Formally,
each connected component of the preimage $O^{-1}(s)$ must be dense, ie
have nonzero measure, in the space of descriptions.

Turing machines in general do not have this property of robustness
against errors. Single bit errors in the input typically lead to
wildly different outcomes. However, an artificial neural network, which is a
computational model inspired by the brain does exhibit this robustness
--- leading to applications such as classifying images in the presence
of noisy or extraneous data.

So what are the chances of the laws of physics breaking down, and of
us finding ourselves in one of Lewis Carrol's creations? Such a
universe will have a very complex description --- for instance the
coalescing of air molecules to form a fire breathing dragon would
involve the complete specification of the states of some $10^{30}$
molecules, an absolutely stupendous amount of information, compared
with the simple specification of the big bang and the laws of physics
that gave rise to life as we know it. The chance of this happening is
equally remote, via Eq. (\ref{Universal Prior}).

\section{QUANTUM MECHANICS}

In the previous sections, I demonstrate that formal mathematical
systems are the most compressible, and have highest measure amongst
all members of the Schmidhuber ensemble. In this work, I explicitly
assume the validity of the {\em Anthropic Principle}, namely that we
live in a description that is compatible with our own existence. This
is by no means a trivial assumption --- it is entirely possible that
we are inhabiting a virtual reality where the laws of the observed
world needn't be compatible with our existence. However, to date, the
Anthropic Principle has been found to be valid\cite{Barrow-Tipler86}.

In order to derive consequences of the Anthropic Principle, one needs
to have a model of consciousness, or at very least some necessary
properties that conscious observer must exhibit. I will explore the
consequences of just two such properties of consciousness.

The first assumption to be made is that observers will find themselves
embedded in a temporal dimension. A Turing machine requires time to
separate the sequence of states it occupies as it performs a
computation. Universal Turing machines are models of how humans
compute things, so it is possible that all conscious observers are
capable of universal computation. Yet for our present purposes, it is
not necessary to assume observers are capable of universal
computation, merely that observers are embedded in time.

The second assumption, which is related to Marchal's {\em
  computational indeter\-minism}\cite{Marchal01}, is that the simple
mathematical description selected from the Schmidhuber ensemble
describes the evolution of an ensemble of possible experiences. The
actual world experienced by the observer is selected randomly from
this ensemble.  More accurately, for each possible experience, an
observer exists to observe that possibility. Since it is impossible to
distinguish between these observers, the internal experience of that
observer is as though it is chosen randomly from the ensemble of
possibilities. This I call the {\em Projection Postulate}.

The reason for this assumption is that it allows for very complex
experiences to be generated from a very simple process. It is a very
generalised form of Darwinian evolution, which exhibits extreme
simplicity over {\em ex nihilo} creation explanations of life on Earth.
Whilst by no means certain, it does seem that a minimum level of
complexity of the experienced world is needed to support conscious
experience of that world according the the anthropic principle.

This ensemble of possibilities at time $t$ we can denote $\psi(t)$.
Ludwig\cite[D1.1]{Ludwig83} introduces a rather similar concept of
ensemble, which he equivalently calls {\em state} to make contact with
conventional terminology. At this point, nothing has been said of the
mathematical properties of $\psi$. I shall now endeavour to show that
$\psi$ is indeed an element from complex Hilbert space, a fact
normally assumed as an axiom in conventional treatments of Quantum
Mechanics.

The projection postulate can be modeled by a partitioning map
$A:\psi\longrightarrow\{\psi_a,\mu_a\}$, where $a$ indexes the
allowable range of potential observable values corresponding to $A$,
$\psi_a$ is the subensemble satisfying outcome $a$
and $\mu_a$ is the measure associated with $\psi_a$ ($\sum_a\mu_a=1$).

Finally, we assume that the generally accepted axioms of set theory
and probability theory hold. Whilst the properties of sets are well
known, and needn't be repeated here, the Kolmogorov
probability axioms are\cite{Li-Vitanyi97}:
\begin{description}
\item[(A1)] If $A$ and $B$ are events, then so is the {\em intersection}
  $A\cap B$, the {\em union} $A\cup B$ and the {\em difference} $A-B$.
\item[(A2)] The {\em sample space} $S$ is an event, called the {\em
    certain event}, and the {\em empty set} $\emptyset$ is an event,
    called the {\em impossible} event.
\item[(A3)] To each event $E$, $P(E)\in[0,1]$ denotes the {\em
    probability} of that event.
\item[(A4)] $P(S)=1$.
\item[(A5)] If $A\cap B=\emptyset$, then $P(A\cup B)=P(A)+P(B)$.
\item[(A6)] For a decreasing sequence 
$
A_1\supset A_2\supset\cdots\supset A_n\cdots
$
of events with $\bigcap_nA_n=\emptyset$, we have
$\lim_{n\rightarrow\infty}P(A_n)=0$. 
\end{description}

Consider now the projection operator ${\cal
  P}_{\{a\}}:V\longrightarrow V$, acting on a ensemble $\psi\in V$, $V$
being the set of all such ensembles, to produce $\psi_a={\cal
  P}_{\{a\}}\psi$, where $a\in S$ is an outcome of an observation. We
have not at this stage assumed that ${\cal P}_{\{a\}}$ is linear.
Define addition for two distinct outcomes $a$ and $b$ as follows:
\begin{equation}\label{add-op}
{\cal P}_{\{a\}}+{\cal P}_{\{b\}} = {\cal P}_{\{a,b\}},
\end{equation}
from which it follows that
\begin{eqnarray}
{\cal P}_{A\subset S} &=& \sum_{a\in A}{\cal P}_{\{a\}}\\
{\cal P}_{A\cup B} &=& {\cal P}_{A} + {\cal P}_{B} -  {\cal P}_{A\cap
  B} \label{union-theorem} \\
{\cal P}_{A\cap B} &=& {\cal P}_{A}{\cal P}_{B} = {\cal P}_{B}{\cal
  P}_{A}.
\end{eqnarray}
These results extend to continuous sets by replacing the discrete sums
by integration over the sets with uniform measure. Here, as elsewhere,
we use $\Sigma$ to denote sum or integral respectively as the index
variable $a$ is discrete of continuous.

Let the ensemble $\psi\in V\equiv\{{\cal P}_{A}\psi| A\subset S\}$ be a
``reference state'', corresponding to the certain event. It encodes
information about the whole ensemble.  Denote the probability of a set
of outcomes $A\subset S$ by $P_\psi({\cal P}_{A}\psi)$. Clearly
\begin{equation}
P_\psi({\cal P}_S\psi) = P_\psi(\psi) = 1
\end{equation}
by virtue of (A4). Also, by virtue of Eq. (\ref{union-theorem})
and (A4), 
\begin{equation}\label{proto-linearity}
P_\psi(({\cal P}_A+{\cal P}_B)\psi) = P_\psi({\cal P}_A\psi) +
P_\psi({\cal P}_B\psi) \;\;\;\mathrm{if}\,A\cap B=\emptyset.
\end{equation}

Assume that Eq. (\ref{proto-linearity}) also holds for $A\cap
B\neq\emptyset$ and consider the possibility that $A$ and $B$ can be
identical. Eq. (\ref{proto-linearity}) may be written:
\begin{equation}\label{natural-linearity}
P_\psi((a{\cal P}_A+b{\cal P}_B)\psi) = aP_\psi({\cal P}_A\psi) +
bP_\psi({\cal P}_B\psi), \forall a,b\in\naturals.
\end{equation}
Thus, the set $V$ naturally extends by means of the addition operator
defined by Eq. (\ref{add-op}) to include all linear combinations
of observed states, at minimum over the natural numbers. If $A\cap
B\neq\emptyset$, then $P_\psi(({\cal P}_A+{\cal P}_B)\psi)$ may exceed
unity, so clearly $({\cal P}_A+{\cal P}_B)\psi$ is not necessarily a
possible observed outcome. How should we interpret these new
``nonphysical'' states? 


At each moment that an observation is possible, an observer faces a
choice about what observation to make. In the Multiverse, the observer
differentiates into multiple distinct observers, each with its own
measurement basis. In this view, there is no {\em preferred
  basis}\cite{Stapp02}. 

The expression $P_\psi((a{\cal P}_A+b{\cal P}_B)\psi)$ must be the
measure associated with $a$ observers choosing to partition the
ensemble into $\{A,\bar{A}\}$ and observing an outcome in $A$ and $b$
observers choosing to partition the ensemble into $\{B,\bar{B}\}$ and
seeing outcome $B$.  The coefficients $a$ and $b$ must be be drawn
from a measure distribution over the possible choices of measurement.
The most general measure distributions are complex, therefore the
coefficients, in general are complex\cite{Cohn80}. We can comprehend
easily what a positive measure means, but what about complex measures?
What does it mean to have an observer with measure $-1$? It turns out
that these non-positive measures correspond to observers who chose to
examine observables that do not commute with our current observable
$A$. For example if $A$ were the observation of an electron's spin
along the $z$ axis, then the states $|+\rangle + |-\rangle$ and
$|+\rangle-|-\rangle$ give identical outcomes as far as $A$ is
concerned. However, for another observer choosing to observe the spin
along the $x$ axis, the two states have opposite outcomes. This is the
most general way of partitioning the Multiverse amongst observers, and
we expect to observe the most general mathematical structures
compatible with our existence.

The probability function $P$ can be used to define an inner product as
follows. Our reference state $\psi$ can be expressed as a sum over the
projected states $\psi=\sum_{a\in S}{\cal
  P}_{\{a\}}\psi\equiv\sum_{a\in S}\psi_a$. Let $V^*={\cal L}(\psi_a)$
be the linear span of this basis set. Then, $\forall \phi, \xi\in V$,
such that $\phi=\sum_{a\in S}\phi_a\psi_a$ and $\xi=\sum_{a\in
  S}\xi_a\psi_a$, the inner product $\langle\phi,\xi\rangle$ is
defined by
\begin{equation}\label{inner product}
\langle\phi,\xi\rangle = \sum_{a\in S}\phi_a^*\psi_a P_\psi(\psi_a).
\end{equation}
It is straightforward to show that this definition has the usual
properties of an inner product, and that $\psi$ is normalized
($\langle\psi,\psi\rangle=1$). The measures $\mu_a$ are given by
\begin{eqnarray}
\mu_a=P_\psi(\psi_a) &=& \langle\psi_a,\psi_a\rangle \nonumber\\
     &=& \langle\psi,{\cal P}_a\psi\rangle\\
     &=& |\langle\psi,\hat\psi_a\rangle|^2\nonumber,
\end{eqnarray}
where $\hat\psi_a=\psi_a/\sqrt{P_\psi(\psi_a)}$ is normalised. 

Until now, we haven't used axiom (A6). Consider a sequence of sets of
outcomes $A_0\supset A_1\ldots$, and denote by $A\subset A_n\forall n$ the
unique maximal subset (possibly empty), such that
$\bar{A}\bigcap_nA_n=\emptyset$. Then the difference ${\cal
  P}_{A_i}-{\cal P}_A$ is well defined, and so
\begin{eqnarray}
\langle ({\cal  P}_{A_i}-{\cal P}_A)\psi, ({\cal  P}_{A_i}-{\cal
  P}_A)\psi\rangle &=& P_\psi(({\cal  P}_{A_i}-{\cal P}_A)\psi)\nonumber \\
&=& P_\psi(({\cal  P}_{A_i}+{\cal  P}_{\bar{A}}-{\cal  P}_S)\psi) \\
&=& P_\psi({\cal  P}_{A_i\cap\bar{A}}).\nonumber
\end{eqnarray}
By axiom (A6), 
\begin{equation} 
\lim_{n\rightarrow\infty} \langle ({\cal P}_{A_i}-{\cal P}_A)\psi,
({\cal P}_{A_i}-{\cal P}_A)\psi\rangle =  0,   
\end{equation}
so ${\cal P}_{A_i}\psi$ is a Cauchy sequence that converges to ${\cal
  P}_{A}\psi\in V$. Hence $V$ is complete under the inner product
(\ref{inner product}). It follows that $V^*$ is complete also, and is
therefore a {\em Hilbert} space.

The most general form of evolution of $\psi$ in continuous time is
given by:
\begin{equation}\label{gen-schrodinger}
\frac{\rmd\psi}{\rmd t}={\cal H}(\psi).
\end{equation}
Some people may think that discreteness of the world's description (ie
of the Schmidhuber bitstring) must imply a corresponding discreteness
in the dimensions of the world. This is not true. Between any two
points on a continuum, there are an infinite number of points that can
be described by a finite string --- the set of rational numbers being
an obvious, but by no means exhaustive example. Continuous systems may
be made to operate in a discrete way, electronic logic circuits being
an obvious example. For the sake of connection with conventional
quantum mechanics, we will assume that time is continuous. A discrete
time formulation can also be derived, in which case we need a
difference equation instead of Eq. (\ref{gen-schrodinger}). Other
possibilities also exist, such as the rational numbers example
mentioned before. The theory of {\em time scales}\cite{Bohner-Peterson01}
could provide a means of developing these other possibilities.

Axiom (A3) constrains the form of the evolution operator ${\cal
  H}$. Since we suppose that $\psi_a$ is also a solution of Eq.
  \ref{gen-schrodinger} (ie that the act of observation does not change
  the physics of the system), ${\cal H}$ must be linear. The certain
  event must have probability of 1 at all times, so 
\begin{eqnarray}
0 &=& \frac{\rmd P_{\psi(t)}(\psi(t))}{\rmd t}\nonumber \\
  &=& \rmd/\rmd t \langle\psi,\psi\rangle\nonumber\\
  &=& \langle\psi,{\cal H}\psi\rangle + \langle{\cal H}\psi,\psi\rangle\nonumber\\
{\cal H}^\dag &=& -{\cal H},
\end{eqnarray}
i.e. ${\cal H}$ is $i$ times a Hermitian operator. 

\section{Discussion}

A conventional treatment of quantum mechanics (see eg
Shankar\cite{Shankar80}) introduces a set of 4-5 postulates that
appear mysterious. In this paper, I introduce a model of observation
based on the idea of selecting actual observations from an ensemble of
possible observations, and can derive the usual postulates of quantum
mechanics aside from the {\em Correspondence Principle}.\footnote{The
  {\em Correspondence Principle} states that classical state variables
  are represented in the quantum formulation by replacing
  appropriately $x\rightarrow X$ and $p\rightarrow -i\hbar
  d/dx$. Stenger\cite{Stenger} has developed a theory based on
  fundamental symmetries that explains the Correspondence Principle.}
Even the property of {\em linearity} is needed to allow disjoint
observations to take place simultaneously in the universe.
Weinberg\cite{Weinberg89,Weinberg92} experimented with a possible
non-linear generalisation of quantum mechanics, however found great
difficulty in producing a theory that satisfied causality. This is
probably due to the nonlinear terms mixing up the partitioning
$\{\psi_a,\mu_a\}$ over time. It is usually supposed that
causality\cite{Tegmark98}, at least to a certain level of
approximation, is a requirement for a self-aware substructure to
exist.  It is therefore interesting, that relatively mild assumptions
about the nature of SASes, as well as the usual interpretations of
probability and measure theory lead to a linear theory with the
properties we know of as quantum mechanics. Thus we have a reversal of
the usual ontological status between Quantum Mechanics and the {\em
  Many Worlds Interpretation}\cite{DeWitt-Graham73}.

\section*{ACKNOWLEDGMENTS}

I would like to thank the following people from the ``Everything''
email discussion list for many varied and illuminating discussions on
this and related topics: Wei Dai, Hal Finney, Gilles Henri, James
Higgo, George Levy, Alastair Malcolm, Christopher Maloney, Jaques
Mallah, Bruno Marchal and J\"urgen Schmidhuber.

In particular, the solution presented here to the White Rabbit paradox
was developed during an email exchange between myself and Alistair
Malcolm during July 1999, archived on the everything list
(\htmladdnormallink{http://www.escribe.com/science/theory}
{http://www.escribe.com/science/theory}).  Alistair's version of this
solution may be found on his web site at
\linebreak\htmladdnormallink{http://www.physica.freeserve.co.uk/p101.htm}
{http://www.physica.freeserve.co.uk/p101.htm}.

I would also like to thank the anonymous reviewer for suggesting
Ludwig's book\cite{Ludwig83}. Whilst the intuitive justification in
that book is very different, there is a remarkable congruence in the set
of axioms chosen to the ones presented in this paper. 


\end{document}